\documentclass[aps,prb,twocolumn,groupedaddress,amsmath,amssymb]{revtex4}

\usepackage{amssymb}
\usepackage{amsmath}
\usepackage{graphicx}
\usepackage{subfigure}
\usepackage{textcomp}
\usepackage{color}
\usepackage{amsfonts}
\usepackage{bbold}
\usepackage{dsfont}
\usepackage{epsfig}
\usepackage{hyperref}
\newcommand{\arctanh}[1]{\operatorname{arctan}}

\begin{document}

%Title of paper
%\title{Inelastic electron transport through spin coupled atomic chains}
\title{Spin-flip inelastic electron tunneling spectroscopy in atomic chains}

\author{Aaron Hurley, Nadjib Baadji and Stefano Sanvito}
\affiliation{School of Physics and CRANN, Trinity College, Dublin 2, Ireland}

\date{\today}

\begin{abstract}
We present a theoretical study of the spin transport properties of mono-atomic magnetic chains with a focus on the spectroscopical 
features of the $I$-$V$ curve associated to spin-flip processes. Our calculations are based on the $s$-$d$ model for magnetism with 
the electron transport treated at the level of the non-equilibrium Green's function formalism. Inelastic spin-flip scattering processes 
are introduced perturbatively via the first Born approximation and an expression for the associated self-energy is derived. The 
computational method is then applied to describe the $I$-$V$ characteristics and its derivatives of one dimensional chains of Mn 
atoms and the results are then compared to available experimental data. We find a qualitative and quantitative agreement between the 
calculated and the experimental conductance spectra. Significantly we are able to describe the relative intensities of the spin excitation 
features in the $I$-$V$ curve, by means of a careful analysis of the spin transition selection rules associated to the atomic chains. 
\end{abstract}

\pacs{75.47.Jn,73.40.Gk,73.20.-r}

\maketitle

%*********************************************************************
% Introduction
%*********************************************************************
\section{Introduction}

Inelastic electron tunneling spectroscopy (IETS) has become an important tool for investigating the elementary excitations
of nanoscale systems~\cite{Jaklevic1966}. An excitation manifests itself as an abrupt change in the differential conductance
of a two-probe device as the voltage sweeps across the excitation energy. This is the result of the opening of a new inelastic 
transport channel for the electron tunneling. As the energy of the probed excitation sets the critical voltage and the temperature 
range where to observe IETS, it is not surprising that the first experiments were all related to molecular vibrations of relatively 
high energy \cite{Stipe1998}.

More recently the continuous advances in low-temperature scanning tunneling microscopy (STM) have enabled the detection of
excitations of different origin, namely those related to the spin degree of freedom. This different type of spectroscopy is usually named 
spin-flip IETS (SF-IETS). The first measurements of SF-IETS were for single
atoms randomly deposited on surfaces~\cite{Heinrich2004}. However, STM techniques also open the possibility of 
assembling and manipulating entire nano-structures~\cite{Eigler} and of positioning magnetic ions on a surface at a desired 
position. This enables the construction of atomic magnetic nano-structures and the study of the complex magnetic 
excitations, resulting from the exchange interaction between the magnetic ions and the substrate, and also between the magnetic
atoms themselves. Such a novel fabrication capability has produced a surge of experimental studies on the spin excitations 
of magnetic nanostructures. These comprise the investigation of the conductance spectra of individual atomic spins~\cite{Hir1}, of 
Kondo screened magnetic dimers~\cite{Hir2} and of multiple chains of Mn atoms~\cite{Hir3}. Interestingly similar investigations 
have now been extended to molecular chains composed of Co-phthalocyanines \cite{Chen}.

The recent rapid growth in the experimental activity has been matched by an equally fast explosion of theoretical works. A 
general and now standard approach to calculating the conductance spectra of the various possible magnetic nanostructures is 
that of combining a master equation solver for the quantum transport problem with model Hamiltonian describing the magnetic 
interaction \cite{Romeike}. This is an intrinsic many-body approach, which in principle contains all the ingredients needed for 
solving the problem, once the various transfer rates are known. As such it usually requires a large number of parameters to
be predictive. An alternative strategy consists in treating the problem at the single particle level, by using a tunnel Hamiltonian
and/or the standard Tersoff-Hamann description \cite{HT} for the STM current~\cite{Fernandez-Rossier,Fransson,Lorente,Persson}.
This second class of computational scheme appears more amenable to be implemented together with first principles electron
transport methods. However it still remains a hybrid theory, where the dynamic part of the problem needs to be approached
at the model Hamiltonian level, although the level of sophistication may include effects related to current generated non-equilibrium 
spin populations~\cite{Sothmann,Delgado}.

A standard theoretical approach to quantum transport is represented by the non-equilibrium Green's function (NEGF) 
formalism~\cite{Keldysh,Datta}, whose mean-field version can be combined with state of the art electronic structure theories
to produce efficient and predictive algorithms \cite{Rocha,Rungger}. This is to date the only fully quantitative quantum
transport approach capable of scaling to large systems~\cite{DNA}, i.e. it is the only one capable of performing simulations for
materials set of current technological relevance. Importantly for this discussion inelastic contributions to the elastic current can also 
be included within the NEGF formalism. In the case of scattering to phonons the problem is usually treated perturbatively by 
constructing an appropriate self-energy at the level of either the first (1BA) or the self-consistent Born approximation (SCBA)~\cite{Galperin}. 
A similar approach to the case of spin excitations is currently not available. 

There are two main reasons for this gap. On the one hand, the adiabatic separation between the electronic degrees of freedom
and those responsible for the inelastic excitations are well defined in the case of nuclear motion (phonons) but less clear in the case 
of spins, since even extremely localized spins have full electronic origin \cite{Antropov}. On the other hand, it is also unclear whether 
the perturbative approach is valid for spins, i.e. whether a suitable expansion parameter can be found. As such, as far as we know, 
an expression for a self-energy describing inelastic spin-flip events has not been derived so far.

In the present work we undertake this challenge and formulate a perturbative theory of spin-flip spectroscopy based on the NEGF
formalism for electron transport. Our theoretical analysis is based on a tight-binding Hamiltonian for the transport electrons, which 
are locally exchange coupled to quantum spins. This is essentially the $s$-$d$ model for magnetism~\cite{Maria}. We then 
proceed in constructing an appropriate self-energy for the spin-degrees of freedom at the 1BA level (note that for this particular problem 
little differences are introduced by extending the treatment to the SCBA) and use this in the standard NEGF scheme for transport. Our 
methodology is then applied to describing SF-IETS in atoms and atomic chains and results are favorably compared with experiments.
% those obtained from the master equation approach and with experiments. 

\section{interaction Hamiltonian and the scattering self-energy}
\label{sec:s-d hamiltonian and the scattering self-energy}

The typical experimental setup considered in this work is that of an STM measurement, i.e. it comprises an STM tip positioned above 
one of the atoms of a magnetic nanostructure, which in turn is weakly coupled to a metallic substrate across an insulating barrier. 
\begin{figure}[ht]
\centering
\resizebox{\columnwidth}{!}{\includegraphics[width=0.5cm,angle=-90]{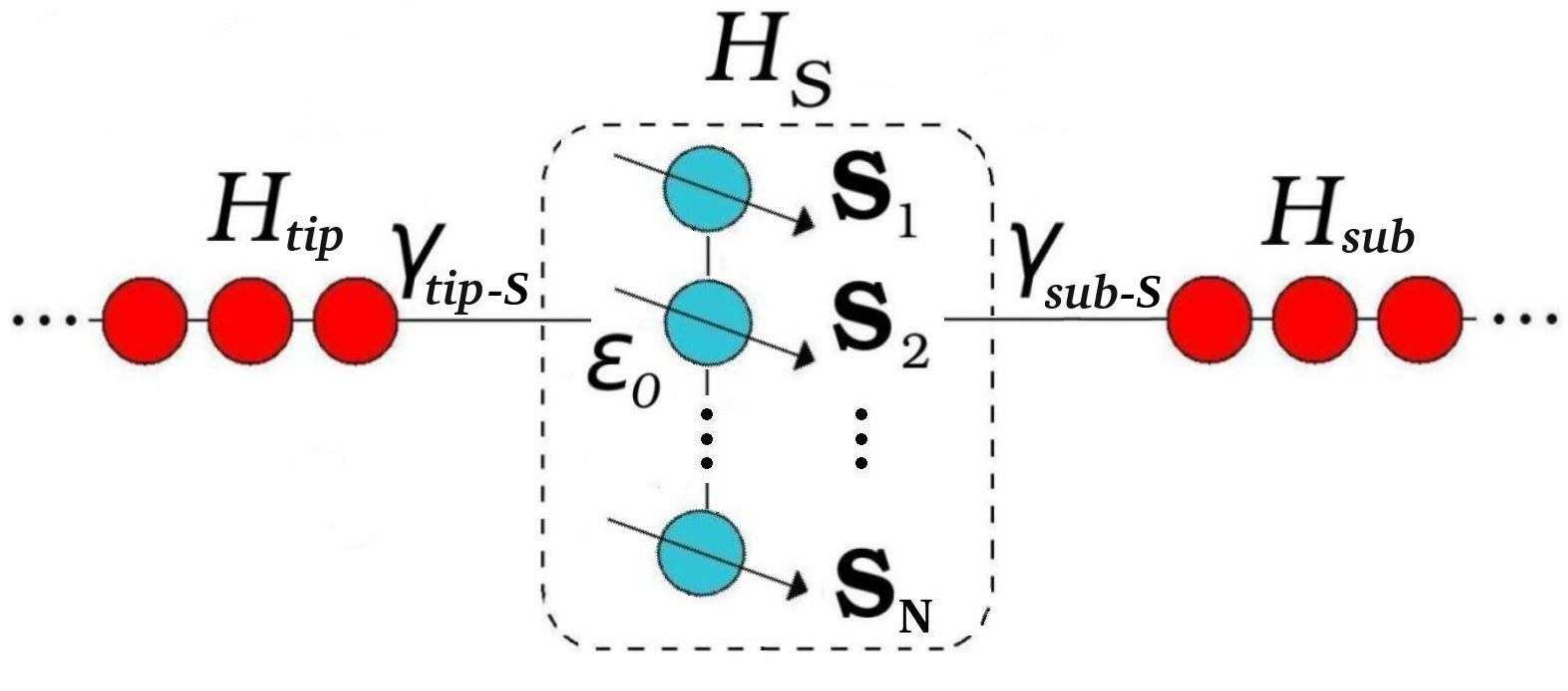}}
\caption{\footnotesize{(Color online) Schematic representation of the device investigated in this work. A scattering region, comprising $N$ 
spin-carrying atoms (light-blue circles) and described by the Hamiltonian $H_\mathrm{S}$, is sandwiched in between two semi-infinite electrodes
(red circles). These mimic the substrate and the tip in a typical STM experiment. The electrodes are non-spin polarized and they are described by 
the Hamiltonian $H_\mathrm{sub}$ and $H_\mathrm{tip}$. In the scattering region the transport electrons are exchange coupled to local 
quantum spins $\mathbf{S}_i$}.}
\label{1}
\end{figure}
We model this system by using a pair of non-interacting semi-infinite leads that are separated by a scattering region as outlined in Fig. \ref{1}. 
The left-hand side lead, the scattering region and the right-hand side lead represent respectively the STM tip, the spin-coupled nanostructure 
and the substrate and they are described by the Hamiltonian $H_\mathrm{tip}$, $H_\mathrm{S}$ and $H_\mathrm{sub}$. For simplicity we
assume an identical electronic structure for both the leads (i.e. they are made of the same material), which we describe by a one orbital per 
site tight-binding Hamiltonian with nearest neighbour interaction. 

The scattering region consists of a one-dimensional chain of $N$ atoms. Each $i$th atom carries a quantum mechanical spin 
$\mathbf{S}_i$ and it is characterized by an on-site energy $\varepsilon_0$. We assume that the tip and substrate can only couple to one 
atom at the time in the scattering region and we describe such a coupling by the hopping integrals $\gamma_\mathrm{tip-S}$ and 
$\gamma_\mathrm{sub-S}$, respectively. This means that the electronic states of the scattering region are broadened by the interaction with the
electrode by $\Gamma_\mathrm{tip-S/sub-S}=\gamma_\mathrm{tip-S/sub-S}^2/\gamma_0$ where $\gamma_0$ is hopping parameter 
in the leads. We assume that $\varepsilon_0\gg\Gamma_\mathrm{t/s}$ leading to a constant density of states at the Fermi energy. The electronic 
effect of the other atoms in the scattering region (i.e. creation of bonding and anti-bonding levels) can be neglected since these levels will also be 
far enough from the Fermi energy to ensure a constant density of states. 

The Hamiltonian of the scattering region contains three parts $H_\mathrm{S}={H}_\mathrm{e}+{H}_\mathrm{sp}+{H}_\mathrm{e-sp}$,
where ${H}_\mathrm{e}$ is the tight-binding electronic part, ${H}_\mathrm{sp}$ is the spin part and ${H}_\mathrm{e-sp}$ describes the 
electron-spin interaction. More explicitly these three components write respectively as
\begin{align}
\label{eq:1}
&{H}_\mathrm{e}=\varepsilon_0\sum_{i\:\alpha}c_{i\alpha}^{\dagger}c_{i\alpha}\:,\\
\label{eq:2}
&{H}_\mathrm{sp}=2J_\mathrm{dd}\sum_{i}^{N-1}\mathbf{S}_i\cdot\mathbf{S}_{i+1}+\\ \nonumber&
+\sum_{i}^N\big\{g{\mu_\mathrm{B}}\mathbf{B}\cdot\bold{S}_i+D({S}^z_i)^2+E[({S}^x_i)^2-({S}^y_i)^2]\big\},\\
\label{eq:3}
&{H}_\mathrm{e-sp}=J_\mathrm{sd}\sum_{i\:\alpha,\alpha'}(c_{i\alpha}^{\dagger}[\boldsymbol{\sigma}_i]_{{\alpha}{\alpha'}}c_{i\alpha'})\cdot\mathbf{S}_i\:.
\end{align}
The electronic part consists only in an on-site potential, i.e. we neglect electron hopping between the sites [the electron ladder operators 
$c_{i\alpha}^{\dagger}/c_{i\alpha}$ create/annihilate an electron at site $i$ with spin $\alpha$ (=$\uparrow,\downarrow$)].

We model the spin-spin interaction between the localized spins $\{\mathbf{S}_i\}$ by a nearest neighbour Heisenberg Hamiltonian with  
coupling strength $J_\mathrm{dd}$. Furthermore we include interaction with an external magnetic field $\mathbf{B}$ ($\mu_\mathrm{B}$ is
the Bohr magneton and $g$ the gyromagnetic ratio) and both uni-axial and transverse anisotropy of magnitude $D$ and  $E$
respectively~\cite{Hir2,Yosida}. The electron-spin interaction Hamiltonian is constructed within the $s$-$d$ model~\cite{Maria} where the 
transport electron, $s$, are locally exchanged coupled to quantum spins, $\{\mathbf{S}_i\}$ ($d$ indicates that the local moments originate 
from the atomic $d$ shell). In this case the interaction strength is $J_\mathrm{sd}$ and in equation (\ref{eq:3}) $\boldsymbol{\sigma}$ is a 
vector of Pauli matrices. 

We now proceed to deriving the perturbation theory in the electron-spin interaction (note that $H_\mathrm{e}$ and $H_\mathrm{sp}$ independently 
can be easily diagonalized). Our strategy is that of first constructing the many-body Green's function at the level of 1BA~\cite{Mahan,Frederiksen,Hyldgaard,Yanik} 
and then, by Dyson's equation, to evaluate the interacting self-energy. In particular we follow closely the procedure laid out in reference~[\onlinecite{Mahan}]. 
Since we consider a non-equilibrium situation at zero temperature our starting point is the Keldysh~\cite{Keldysh} contour-ordered single-body Green's function 
in the many-body ground state,
\begin{equation}
\label{eq:4} [G(\tau,\tau')]_{\sigma\sigma'}=-i{\langle}|T_C\{c_{\sigma}(\tau)c^{\dagger}_{\sigma'}(\tau')\}|{\rangle}\:,
\end{equation}
where the time-average is performed over the full interacting ground state $|{\rangle}$ (note that for clarity we have dropped the site index $i$, which
will be explicitly included only where necessary). Equation (\ref{eq:4}) can be expanded up to the $n$-th order in the interaction Hamiltonian, 
$H_\mathrm{e-sp}$, as
\begin{align}
\label{eq:5}
&[G(\tau,\tau')]_{\sigma\sigma'}=\sum_n\frac{(-i)^{n+1}}{n!}\int\limits_C{d}\tau_1\dots\int\limits_C{d}\tau_n\ \times \nonumber \\ &\frac{{\langle}0|T_C\{{H}_\mathrm{e-sp}(\tau_1)\dots{H}_\mathrm{e-sp}(\tau_n)c_{k\sigma}(\tau)c_{k'\sigma'}^{\dagger}(\tau')\}|0{\rangle}}{U(-\infty,-\infty)},
\end{align}
where $U$ is the time-evolution unitary operator and the time-averages are now over the known non-interacting $(J_\mathrm{sd}=0)$ ground state 
$|0{\rangle}$. The time integration over $\tau$ is ordered on the contour $C$ going from $-\infty$  to $+\infty$ and then returning from $+\infty$ to
$-\infty$, since the ground state of the non-equilibrium system can only be defined at $-\infty$\cite{Haug}. If the expansion is truncated to the 
first order one obtains a Zeeman-like term which can be neglected as long as $\varepsilon_0\gg\gamma_\mathrm{tip-S/sub-S}$. 
The first contribution of interest then appears at the second order. This can be obtained by inserting the explicit expression for ${H}_\mathrm{e-sp}(t)$ 
[Eq. (\ref{eq:3})] into Eq.~(\ref{eq:4}) 
\begin{align}
\label{eq:6}
&[G(\tau,\tau')]^{(2)}_{\sigma\sigma'}=\frac{(-i)^{3}}{2!}J^2_\mathrm{sd}\sum_{i,\alpha,\alpha',j,\beta,\beta'}\int\limits_C{d}\tau_1\int\limits_C{d}\tau_2\ \times \nonumber \\ &{\langle}0|T_C\{c_{\sigma}(\tau)c_{\alpha}^{\dagger}(\tau_1)c_{\alpha'}(\tau_1)c_{\beta}^{\dagger}(\tau_2)c_{\beta'}(\tau_2)c_{\sigma'}^{\dagger}(\tau')\}|0{\rangle} \nonumber \\
&\times {\langle}0|T_C\{{S^{i}}(\tau_1){S^{j}}(\tau_2)\}|0{\rangle}[{\sigma}^i]_{{\alpha}{\alpha'}}[{\sigma}^j]_{{\beta}{\beta'}},
\end{align}
where the indices $i$ and $j$ now run over the cartesian coordinates $x$, $y$ and $z$. 

A full contour-ordered expansion must now be performed on both the electron bracket and the spin bracket. The electron bracket has six different 
time-ordering combinations, which are explicitly listed in the equation~(\ref{eq:7}) below
\begin{align}
\label{eq:7}
{\langle}0|T_C\{c_{\sigma}(\tau)c_{\alpha}^{\dagger}(\tau_1)c_{\alpha'}(\tau_1)c_{\beta}^{\dagger}(\tau_2)c_{\beta'}(\tau_2)c_{\sigma'}^{\dagger}(\tau')\}|0{\rangle}&\ \ \ \ \ \ \ \nonumber \\
={\langle}0|T_C\{c_{\sigma}(\tau)c_{\alpha}^{\dagger}(\tau_1)\}|0{\rangle}\times {\langle}0|T_C\{c_{\alpha'}(\tau_1)c_{\beta}^{\dagger}(\tau_2)\}|0{\rangle}\nonumber \\ \times 
{\langle}0|T_C\{c_{\beta'}(\tau_2)c_{\sigma'}^{\dagger}(\tau')\}|0{\rangle} \nonumber \\
+{\langle}0|T_C\{c_{\sigma}(\tau)c_{\beta}^{\dagger}(\tau_2)\}|0{\rangle}\times {\langle}0|T_C\{c_{\alpha'}(\tau_1)c_{\sigma'}^{\dagger}(\tau')\}|0{\rangle}\nonumber \\ \times  {\langle}0|T_C\{c_{\alpha}(\tau_2)c_{\beta'}^{\dagger}(\tau_1)\}|0{\rangle} \nonumber \\
+{\langle}0|T_C\{c_{\sigma}(\tau)c_{\alpha}^{\dagger}(\tau_1)\}|0{\rangle}\times {\langle}0|T_C\{c_{\alpha'}(\tau_1)c_{\sigma'}^{\dagger}(\tau')\}|0{\rangle}\nonumber \\ \times  {\langle}0|T_C\{c_{\beta}^{\dagger}(\tau_2)c_{\beta'}(\tau_2)\}|0{\rangle} \nonumber \\
+{\langle}0|T_C\{c_{\sigma}(\tau)c_{\beta}^{\dagger}(\tau_2)\}|0{\rangle}\times {\langle}0|T_C\{c_{\beta'}(\tau_2)c_{\sigma'}^{\dagger}(\tau')\}|0{\rangle}\nonumber \\ \times  {\langle}0|T_C\{c_{\alpha}^{\dagger}(\tau_1)c_{\alpha'}(\tau_1)\}|0{\rangle} \nonumber \\ 
+{\langle}0|T_C\{c_{\sigma}(\tau)c_{\sigma'}^{\dagger}(\tau')\}|0{\rangle}\times {\langle}0|T_C\{c_{\alpha}^{\dagger}(\tau_1)c_{\alpha'}(\tau_1)\}|0{\rangle}\nonumber \\ \times  {\langle}0|T_C\{c_{\beta}^{\dagger}(\tau_2)c_{\beta'}(\tau_2)\}|0{\rangle} \nonumber \\
-{\langle}0|T_C\{c_{\sigma}(\tau)c_{\sigma'}^{\dagger}(\tau')\}|0{\rangle}\times {\langle}0|T_C\{c_{\alpha'}(\tau_1)c_{\beta}^{\dagger}(\tau_2)\}|0{\rangle}\nonumber \\ \times  {\langle}0|T_C\{c_{\beta'}(\tau_2)c_{\alpha}^{\dagger}(\tau_1)\}|0{\rangle}.
\end{align}
The last two combinations can be eliminated since they represent unconnected Feynman diagrams which vanish in the averaging 
process~\cite{Mahan}. The first and the second term are equal under the exchange of the indexes, and so are the third and 
the fourth ones. This leaves us with a simplified expression which, when compared to Eq. (\ref{eq:5}), gives us
\begin{align}
\label{eq:8}
&{\langle}0|T_C\{c_{\sigma}(\tau)c_{\alpha}^{\dagger}(\tau_1)c_{\alpha'}(\tau_1)c_{\beta}^{\dagger}(\tau_2)c_{\beta'}(\tau_2)c_{\sigma'}^{\dagger}(\tau')\}|0{\rangle}\nonumber \\
&=2i^3\delta_{\sigma\alpha}\delta_{\alpha'\beta}\delta_{\beta'\sigma'}[G_0(\tau,\tau_1)]_{\sigma\sigma}[G_0(\tau_1,\tau_2)]_{\alpha'\alpha'}\nonumber \\ &\ \ \ \ \ \ \ \ \ \ \ \ \ \ \ \ \ \ \ \ \ \ \ \ \ \ \ \ \
\times[G_0(\tau_2,\tau')]_{\sigma'\sigma'} \nonumber \\
&+\ 2i^3\delta_{\sigma\alpha}\delta_{\alpha'\sigma'}\delta_{\beta\beta'}[G_0(\tau,\tau_1)]_{\sigma\sigma}[G_0(\tau_1,\tau')]_{\sigma'\sigma'}\nonumber \\ &\ \ \ \ \ \ \ \ \ \ \ \ \ \ \ \ \ \ \ \ \ \ \ \ \ \ \ \ \
\times[G_0(\tau_2,\tau_2)]_{\beta\beta}.
\end{align}
In this case, since the averaging bracket is over the non-interacting ground state, $G_0$ represents the un-perturbed electronic Green's function and can 
be calculated exactly.

We then return to Eq.~(\ref{eq:6}) and evaluate the spin bracket. The ground state of the non-interacting spin system alone can be found by diagonalizing 
exactly ${H}_\mathrm{sp}$. This is achieved by constructing the full basis $\{|n{\rangle}\}$ where $n=-S,-S+1,...,+S$. The resulting eigenvectors, $|m{\rangle}$, 
and eigenvalues, $E_m$, satisfy the Schr\"odinger equation ${H}_\mathrm{sp}|m{\rangle}=\varepsilon_m|m{\rangle}$ and they can be used to re-write the operators $S^i(\tau)$ 
as
\begin{align}
\label{eq:9}
S^i(\tau)=\sum_{m,n}{\langle}m|S^i|n{\rangle}d^{\dagger}_m(\tau)d_n(\tau)\:.
\end{align}
Here $d_n$ is a desctruction operator for a quasi-particle of the spin system. The quasi-particles are then assumed to be fermionic in nature so that they obey the 
anticommutation rules $[d_m^{\dagger},d_n]=\delta_{mn}$ and $[d_m^{\dagger},d_n^{\dagger}]=[d_m,d_n]=0$. Such an assumption is valid as long as the excitations 
considered are always around the ground state, i.e. under the condition that the spin system can always efficiently relax back to the ground state between spin-flip events. 
We can then define a contour-ordered spin Green's function as follows
\begin{align}
\label{eq:10}
[D(\tau,\tau')]_{n,m}=-i{\langle}|T_C\{d_{n}(\tau)d^{\dagger}_{m}(\tau')\}|{\rangle}\:.
\end{align}
By inserting the expressions in the equations (\ref{eq:9}) and (\ref{eq:10}) into the spin bracket and by computing the time-ordered contraction 
we finally obtain
\begin{align}
\label{eq:11}
&{\langle}0|T_C\{{S_{i}^{+}}(\tau_1){S_{j}^{-}}(\tau_2)\}|0{\rangle}= \nonumber \\ 
&-\sum_{m,n}{\langle}m|S^i|n{\rangle}{\langle}n|S^j|m{\rangle}[D_0(\tau_1,\tau_2)]_{n,n}[D_0(\tau_2,\tau_1)]_{m,m}\:.
\end{align}

The set of equations (\ref{eq:8}) and (\ref{eq:11}) can now be incorporated into the expression for the second order contribution to the 
many-body Green's function [Eq.~(\ref{eq:6})]. Then, by using Dyson's equation~\cite{Mahan}, one can finally write the second order 
contribution to the interacting self-energy which reads
\begin{align}
\label{eq:12}
[\Sigma_\mathrm{int}(&\tau_1,\tau_2)]^{(2)}_{\sigma\sigma'}=-J^2_\mathrm{sd}\sum_{i,j,\beta}\Big\{[{\sigma}^i]_{{\sigma}{\beta}}[{\sigma}^j]_{{\beta}{\sigma'}}+[{\sigma}^i]_{\sigma\sigma'}[{\sigma}^j]_{\beta\beta}\Big\}\nonumber \\
&\times[G_0(\tau_1,\tau_2)]_{\beta\beta}\sum_{m,n}{\langle}m|S^i|n{\rangle}{\langle}n|S^j|m{\rangle} \nonumber \\
&\times[D_0(\tau_1,\tau_2)]_{n,n}[D_0(\tau_2,\tau_1)]_{m,m}\:.
\end{align}
If we now assume that the ground state electronic spin levels are degenerate, i.e. $[G_0]_{\uparrow\uparrow}=[G_0]_{\downarrow\downarrow}$, 
then the only quantity of interest is the trace of the self-energy over spin indices. By performing such a trace, the spin-independent self-energy finally reads
\begin{align}
\label{eq:13}
\Sigma_\mathrm{int}(&\tau_1,\tau_2)^{(2)}=-2J^2_\mathrm{sd}\sum_{i,m,n}|{\langle}m|S^i|n{\rangle}|^2G_0(\tau_1,\tau_2)\nonumber \\
&\times[D_0(\tau_1,\tau_2)]_{n,n}[D_0(\tau_2,\tau_1)]_{m,m}\:.
\end{align}
where we have used the results $\text{Tr}[\sigma^i\sigma^j]=\delta_{ij}$ and $\text{Tr}[\sigma^i]=0$. 

At this point we can calculate the real-time quantities, such as the lesser (greater) self-energies, by using the Langreth's theorem for time ordering over the 
contour, $\tau_1\in C_1(C_2)$ and $\tau_2\in C_2(C_1)$ \cite{Haug}. $C_1$ is the time ordering contour from $-\infty$ to $+\infty$ and $C_2$ is the time 
anti-ordering contour from $+\infty$ to $-\infty$. We find
\begin{align}
\label{eq:14}
&\Sigma^{\lessgtr}_\mathrm{int}(t_1,t_2)^{(2)}=-2J^2_\mathrm{sd}\sum_{i,m,n}|{\langle}m|S^i|n{\rangle}|^2G^{\lessgtr}_0(t_1,t_2)\nonumber \\ & \ \ \ \ \ 
\times[D^{\lessgtr}_0(t_1,t_2)]_{n,n}[D^{\gtrless}_0(t_2,t_1)]_{m,m}\nonumber \\
&\ \ \ \ \ \ \ \ \ \ \ \ \ \ \ \ \ \  =-2J^2_\mathrm{sd}\sum_{i,m,n}|{\langle}m|S^i|n{\rangle}|^2G^{\lessgtr}_0(t_1,t_2)\nonumber \\& \ \ \ \ \ \times P_n(1-P_m)e^{\pm i(\varepsilon_m-\varepsilon_n)(t_1-t_2)}\:,
\end{align}
where in the second step we have written explicitly $D^{\lessgtr}_0(t_1,t_2)$ in terms of the spin-level occupations, $P_n=d^{\dagger}_nd_n$. The dependence
of $\Sigma^{\lessgtr}_\mathrm{int}$ over the energy, $E$, can found by simple Fourier transform
\begin{align}
\label{eq:15}
&\Sigma^{\lessgtr}_\mathrm{int}(E)^{(2)}=-2J^2_\mathrm{sd}\sum_{i,m,n}|{\langle}m|S^i|n{\rangle}|^2P_n(1-P_m)\nonumber \\
&\ \ \ \ \ \times G^{\lessgtr}_0(E\pm\Omega_{mn})\:,
\end{align}
where $\Omega_{mn}=\varepsilon_m-\varepsilon_n$ and where the $+$ ($-$) symbol corresponds to $\Sigma^{<}$ ($\Sigma^{>}$).

\begin{figure}[h]
\centering
\resizebox{\columnwidth}{!}{\includegraphics[,angle=-90]{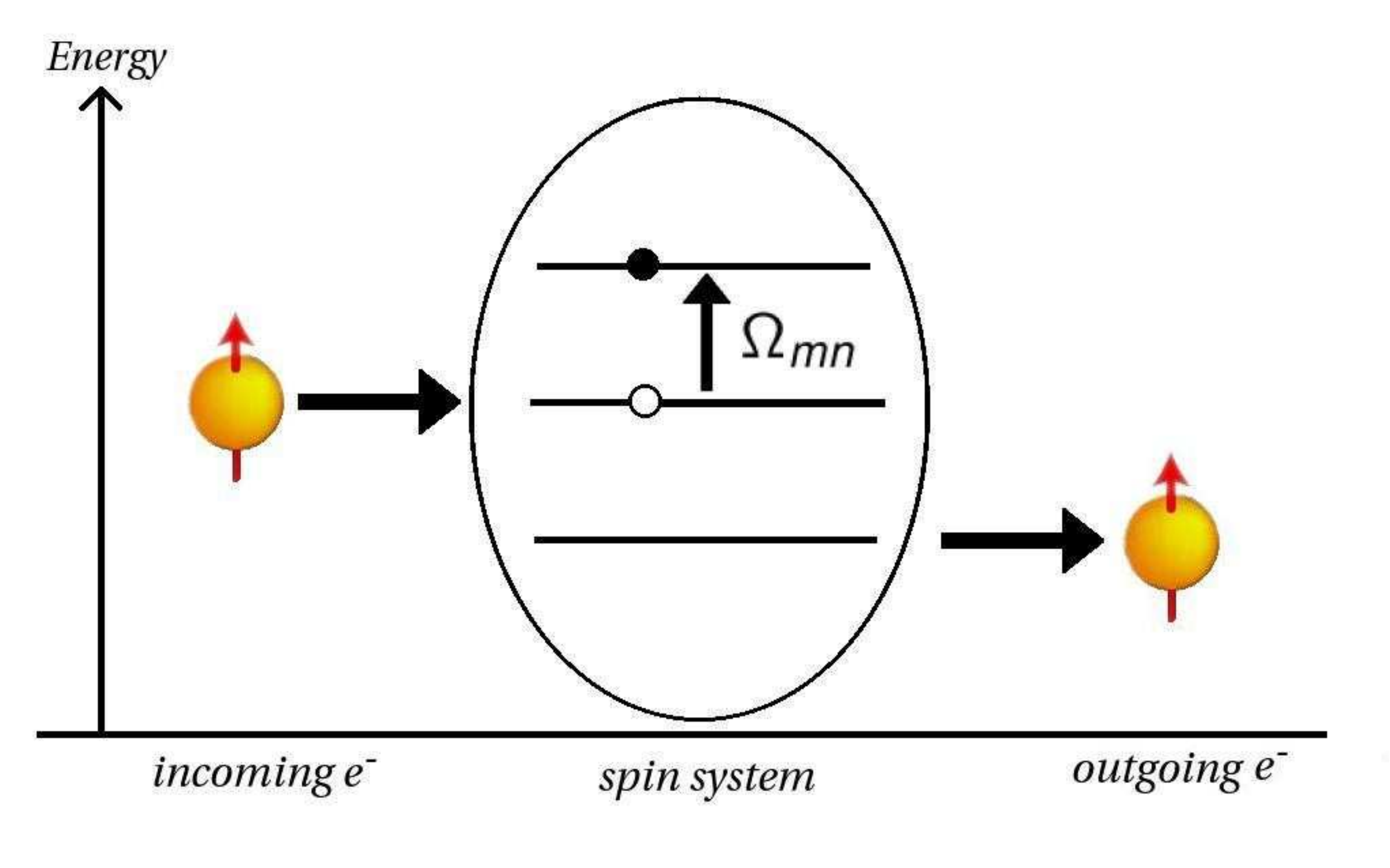}}
\caption{\footnotesize{Schematic representation of the inelastic process described by the greater self-energy, $\Sigma^{>}_\mathrm{int}$. An incoming electron 
scatters against a localized spin and decreases its energy by $\Omega_{mn}$. This is transfer to the local spin system, which undergos a spin transition
$|n\rangle\rightarrow|m\rangle$.}}
\label{2}7
\end{figure}

Electron-spin scattering events are now fully described by $[\Sigma_{int}^{\lessgtr}(E)]$. In particular Eq.~(\ref{eq:15}) describes the process where 
an incoming electron with energy $E$ experiences a spin-flip process, which changes its energy by $\pm\Omega_{mn}$. This is the result of the 
electron-spin interaction with the local spins. Such a process is schematically represented in Fig.~\ref{2}. Note that the probability for an excitation
to occur is determined by the prefactors ${\langle}m|S^i|n{\rangle}|^2P_n(1-P_m)$, i.e. by the state of the spin system.

\section{non-equilibrium green's function method for electron transport}

\label{sec:NON-EQUILIBRIUM GREEN'S FUNCTION (NEGF) METHOD}

The transport properties of the device are described by using the non-equilibrium Green's function (NEGF) method. This has been extensively 
described in the past~\cite{Keldysh,Datta,Rocha}. Here we only summarize the main concepts and we highlight the modifications associated to 
describing electron-spin interaction. A two-probe device can be divided into three distinct regions, two semi-infinite leads which represent the STM 
tip and the underlying substrate and a central scattering region (see Fig.~\ref{1}). The leads act as charge reservoirs and they are characterized 
by their chemical potential, respectively $\mu_\mathrm{tip}$ and $\mu_\mathrm{sub}$. The external bias is introduced in the form of a relative shift 
(symmetric) of the two chemical potentials with respect to each other. The underline assumption of the method is that under the external bias there 
is no rearrangement of the electronic structure of the leads, i.e. that electron screening in the leads is efficient. This simplifies the problem to that of 
calculating the retarded Green's function of the scattering region \cite{Datta,Yanik,Rocha} only
\begin{equation}
\label{eq:17}
G(E)=\lim_{\eta \to 0}[(E-i\eta)I-H_\mathrm{e}-\Sigma(E)]^{-1}\:.
\end{equation}
Here $H_\mathrm{e}$ is the electronic part of the Hamiltonian [Eq.~(\ref{eq:1})] and $\Sigma(E)$ is the retarded self-energy, which incorporates 
the effects of the leads and of the inelastic interaction. In particular $\Sigma(E)$ writes as
\begin{equation}
\label{eq:18}
\Sigma(E) = \Sigma_\mathrm{tip}(E)+\Sigma_\mathrm{sub}(E)+\Sigma_\mathrm{int}(E)\:,
\end{equation}
where $\Sigma_\mathrm{tip}(E)$ and $\Sigma_\mathrm{sub}(E)$ are respectively the STM tip and substrate self-energies, while $\Sigma_\mathrm{int}$ 
is the scattering self-energy describing the electron-spin interaction. The leads' self-energies can be written in terms of the surface Green's functions 
$\hat{g}_\mathrm{tip}$ and $\hat{g}_\mathrm{sub}$ and the coupling matrices between the leads and the scattering region $H_{\alpha-\mathrm{S}}$ 
($\alpha$=tip, sub)
\begin{align}
\label{eq:19}
\Sigma_\mathrm{tip}(E)=H^{\dagger}_\mathrm{tip-S}\:\hat{g}_\mathrm{tip}(E)\:H_\mathrm{tip-S}\:,\\
\label{eq:20}
\Sigma_\mathrm{sub}(E)=H_\mathrm{sub-S}\:\hat{g}_\mathrm{sub}(E)\:H^{\dagger}_\mathrm{sub-S}\:.
\end{align}

The surface Green's functions can be found by first constructing the Green's function for an infinite system and then by applying the appropriate boundary 
conditions~\cite{Sanvito}. For a single-site nearest-neighbour one-dimensional tight-binding chain $\hat{g}_\alpha$ takes a simple close form
\begin{align}
\label{eq:21}
\hat{g}_\mathrm{tip}(E)&= \hat{g}_\mathrm{sub}(E)=\frac{1}{\gamma_0}e^{ik(E)}\:,\\
\label{eq:22}
k(E)&=\cos^{-1}\left(\frac{E-\varepsilon_0}{2\gamma_0}\right),
\end{align}
where we $k(E)$ and $\gamma_0$ are respectively the dispersion relation and the hopping parameter. We assume equal hopping parameter in the leads 
and also the condition $\gamma_0>\varepsilon_0$. The interaction between the scattering region and the leads has the effect of broadening the scattering 
region discrete energy levels. The broadening functions write
\begin{equation}
\label{eq:23}
\Gamma_{\alpha}(E)=i[\Sigma_{\alpha}(E)-\Sigma_{\alpha}^{\dagger}(E)]\:,
\end{equation}
where ($\alpha$=tip, sub). The resulting lesser/greater leads self-energies are related to the broadening and to the population of the 
electrons/holes in the leads
\begin{align}
\label{eq:24}
\Sigma_{\alpha}^{<}(E)&=f_{\alpha}(E)\Gamma_{\alpha}(E),\\
\label{eq:25}
\Sigma_{\alpha}^{>}(E)&=[1-f_{\alpha}(E)]\Gamma_{\alpha}(E),
\end{align}
where $f_{\alpha}(E)=1/\left[1+\exp\left(\frac{E-\mu_\alpha}{k_BT}\right)\right]$ are the Fermi functions.

Finally, the retarded electron-spin scattering self-energy is found from the Hilbert transform\cite{Yanik}
\begin{align}
\label{eq:26}
\Sigma_\mathrm{int}(E)=\mathcal{PV}\int\limits_{-\infty}^{\infty}\frac{{d}E'}{2\pi}\frac{\Sigma_\mathrm{int}^>(E')-\Sigma_\mathrm{int}^<(E')}{E'-E}+\nonumber \\ 
-\:i\frac{\Sigma_\mathrm{int}^>(E)-\Sigma_\mathrm{int}^<(E)}{2},
\end{align}
where $\mathcal{PV}$ denotes the principal value and where the full expressions for $\Sigma_\mathrm{int}^>(E)$ and $\Sigma_\mathrm{int}^<(E)$ 
have been given in the previous section. Note that at variance with a similar expression for the self-energy describing electron-phonon scattering,
in equation~(\ref{eq:26}) there is no Hartree-like contribution. This usually describes polaronic effects and it is known to cause the breakdown of the 
Born approximation for strong electron-phonon coupling~\cite{Will}.

Additional quantities of interest, which enter in the definition of the electron current, are the lesser and greater Green's functions ($G^<$ and $G^>$). Their 
diagonal elements respectively give the electron and hole density in the scattering region. These are defined by
\begin{equation}
\label{eq:27}
G^{\lessgtr}(E)=G(E)[\Sigma^{\lessgtr}(E)]G^\dagger(E)\:,
\end{equation}
where 
\begin{equation}
\label{eq:28}
\Sigma^{\lessgtr}(E)=\Sigma^{\lessgtr}_\mathrm{tip}(E)+\Sigma^{\lessgtr}_\mathrm{sub}(E)+\Sigma^{\lessgtr}_\mathrm{int}(E)\:.
\end{equation}
As already mention the application of a potential difference $V$ across the device produces a shift in the tip and substrate chemical potentials.
By assuming, as done here, that the tip and the substrate share the same Fermi level $E_\mathrm{F}$ then we have $\mu_\mathrm{tip}=E_\mathrm{F}+eV/2$
and $\mu_\mathrm{sub}=E_\mathrm{F}+eV/2$. This is equivalent to replacing the energy in the leads with $E \rightarrow E \pm eV/2$ ($e$ is the electron
charge). Finally the current can then be calculated for a range of values of $V$ at any of the leads $\alpha$ as
\begin{align}
\label{eq:29}
&I_\alpha=\int_{-\infty}^{+\infty}\bar{I}_{\alpha}(E)\:\mathrm{d}E\:,\\
\label{eq:30}
\bar{I}_{\alpha}(E)=\frac{e}{h}\mathrm{Tr}\{&[\Sigma_{\alpha}^{>}(E)G^{<}(E)]-[\Sigma_{\alpha}^{<}(E)G^{>}(E)]\}\:.
\end{align}

%%%%%%%%%%%%%%%%%%%%%%%%%%%%%%%%%%%%%%%%%%%%%%%%%

\section{Results}
\label{sec:Results}

The entire procedure outlined in the previous two sections is now implemented numerically in order to calculate the $I$-$V$ characteristics 
and the conductance spectra ($\mathrm{d}I/\mathrm{d}V$-$V$) of some selected magnetic nanostructures. In particular we aim at reproducing 
the $\mathrm{d}I/\mathrm{d}V$-$V$ curve observed through Mn monoatomic chains deposited on thin CuN insulating atomic layers, as 
described in reference [\onlinecite{Hir3}]. Clearly our 1D model does not include all the details involved in the actual SF-IETS STM experiments, 
but we argue here that it contains already all the ingredients to reproduce the main experimental features. 

In Table I we list all the parameters used in our simulations and their assigned values. These have been either inferred from the experiments 
(Exp)~\cite{Hir3} or have been estimated from density functional theory (DFT) calculations~\cite{Zitko}.
\begin{table}[t]
\centering                          % used for centering table
\begin{tabular}{c c c c}            % centered columns (4 columns)
\hline\hline                        %inserts double horizontal lines
Quantity & Symbol & Value & Origin [Ref.] \\ [0.5ex]   % inserts table
%heading
\hline                              % inserts single horizontal line
Atomic Apin & $S$ & $\frac{5}{2}$ & Exp. [\onlinecite{Hir3}]  \\               % inserting body of the table
Temperature & $T$ & 0.6 K & Exp. [\onlinecite{Hir3}] \\
{d-d} exchange & $J_\mathrm{dd}$ & +6.2 meV & Exp. [\onlinecite{Hir3}] \\
{s-d} exchange  & $J_\mathrm{sd}$ & +500 meV & DFT [\onlinecite{Zitko}] \\
Fermi Energy & $E_\mathrm{F}$ & 0 meV & DFT [\onlinecite{Zitko}] \\% [1ex] adds vertical space
Lead hopping integral & $\gamma_0$ & 10000 meV & DFT [\onlinecite{Zitko}] \\ 
Channel on-site energy & $\varepsilon_0$ & 1000 meV & DFT [\onlinecite{Zitko}] \\ 
Tip-channel hopping & $\gamma_\mathrm{tip-S}$ & 50 meV & DFT [\onlinecite{Zitko}] \\
Substrate-channel hopping & $\gamma_\mathrm{sub-S}$ & 500 meV & DFT [\onlinecite{Zitko}] \\  
Axial anisotropy & $D$ & -0.037 meV & Exp. [\onlinecite{Hir1}] \\ 
Transverse anisotropy & $E$ & 0.007 meV & Exp. [\onlinecite{Hir1}]\\ [1ex]
\hline                              %inserts single line
\end{tabular}
\label{table1}          % is used to refer this table in the text
\caption{\footnotesize{Empirical parameters used in the numerical simulations presented in this work and their assigned numerical values. 
In the rightmost column we report the source of a particular estimate. ``Channel'' here means the scattering region.}}
\end{table}
The local Mn spin is set to be $5/2$, as proposed in the original experimental works~\cite{Hir1,Hir3}, confirmed by DFT calculations~\cite{Zitko}
and expected from the nominal Mn valence. The spin-spin coupling parameter $J_\mathrm{dd}$ corresponds to an antiferromagnetic order 
between the neighbouring Mn spins, a feature verified in the experimental conductance spectra. The lead on-site energy is suitably set to 
zero and simply defines the reference potential. We also note that the value for $J_\mathrm{sd}$ is determined from theory to be of the order 
of 500meV~\cite{Lucignano}. Although we employ a zero-temperature formalism we nevertheless evaluate the Fermi functions of the leads at 
the small temperature of 0.6~K. This allows us to include minor thermal smearing of the electron gas in the leads and consequently of the 
conductance profile. Finally we notice that the scattering region is expected to be significantly more strongly coupled to the substrate than 
to the tip. This essentially means that the spin population in the scattering region remains effectively in equilibrium~\cite{Sothmann} with 
the substrate, i.e. there is no accumulation of spin excitations between two inelastic tunneling events.

\begin{figure}[h]
\centering
\resizebox{\columnwidth}{!}{\includegraphics[width=2cm,angle=-90]{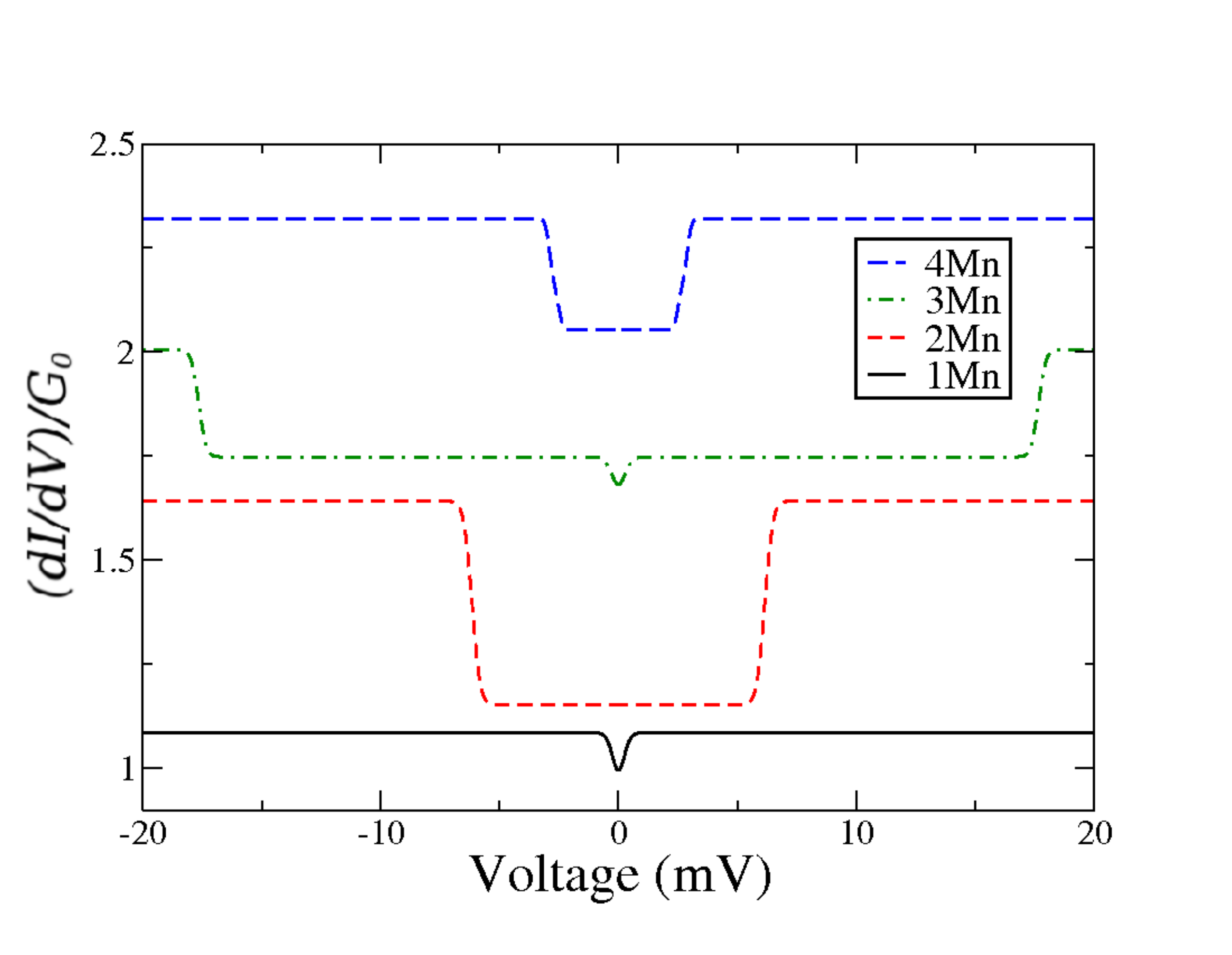}}
\caption{\footnotesize{(Color online) Normalised conductance spectra for Mn chains of different lengths $N$. The various spectra are offset for clarity. 
The tip is placed above the second atom of the chain for chains with $N>2$. We notice the strong dependence of the spectra on the chain parity.}}
\label{3}
\end{figure}
Figure~\ref{3} shows the calculated conductance spectra (normalised against the elastic contribution to the conductance $G_0$) for $N$-atom 
long Mn chains ($N\le4$) in no external magnetic field. In the case of $N$=3 and $N$=4 the spectrum is calculated for the STM tip placed above 
the second atom in the chain. From the figure it is clear that all the qualitative experimental findings are correctly reproduced. In particular we 
notice the strong dependence of the conductance profile over the parity of the chains, with chains comprising an odd number of atoms (odd chains) 
exhibiting a conductance dip at around $V$=$0$, which is absent for even chains.
 
The ground state of each odd chain has a net total spin $S_\mathrm{tot}=5/2$. This is affected by the transverse and axial anisotropy, which 
lift the ground state degeneracy and allow a transition between the ground state and the first spin-flip excited state to occur. The excitation of such 
a transition results in a conductance step at a voltage corresponding to the transition energy. Since the anisotropy energies are small (see Table I), 
the excitation energy is small as well and the feature in the conductance profile appears near $V$=0. In contrast the even chains do not show any 
zero-bias excitation, since the ground state is a singlet and does not carry a magnetic moment. This is a direct proof that the magnetic interaction 
between Mn ions is antiferromagnetic. Should this have been ferromagnetic, even chains would also have shown zero-bias anomalies. 

\begin{figure}[h]
\centering
\resizebox{\columnwidth}{!}{\includegraphics[width=5cm,angle=-90]{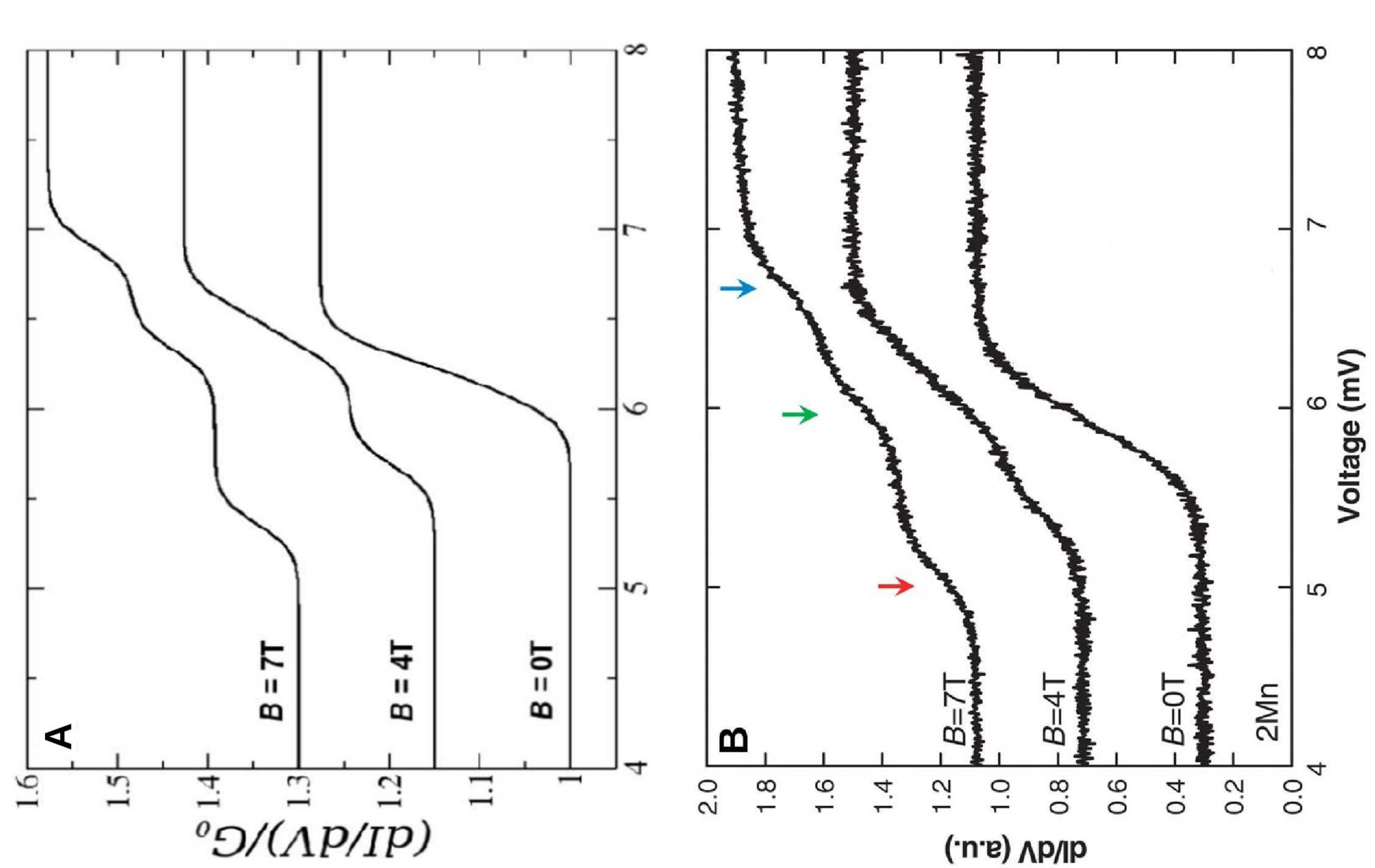}}
\caption{\footnotesize{(Color online) Theoretical (A) and experimental (B) conductance spectra for the $N$=2 chains in a finite external magnetic 
field. For $\mathbf{B}\ne0$ we note a splitting of the conductance steps, corresponding to the Zeeman split of the final triplet excited state. This is 
observed in the experiments of reference [\onlinecite{Hir3}], which are also reported in panel (B).}}
\label{4}
\end{figure}

As mentioned above the conductance steps encountered at around $V$=0 correspond to spin-flip events, i.e. to electron scattering processes
that produce the transition $|m=5/2\rangle\rightarrow|m=3/2\rangle$ ($m$ is the magnetic quantum number) but that also preserve the total 
spin $S_\mathrm{tot}=5/2$. The first net spin changing transition occurs for $N=2$ and corresponds to the large conductance step found at 
$V=J_\mathrm{dd}/e=6.2$~mV. This is investigated in more detail in figure~\ref{4}, where we also include the dimer's spectrum calculated
when a magnetic field is applied along the $z$-direction. The effect of the magnetic field is that of splitting the single excitation line into
three distinct conductance steps. In this case in fact the antiferromagnetic dimer has a singlet ($S_\mathrm{tot}$=0) ground state and a
triplet ($S_\mathrm{tot}$=1) first excited state. Therefore, an excitation from the ground state to the first excited state changes the net spin 
of the dimer. When a magnetic field is applied the degeneracy of the triplet excited state is lifted and excitations having the three 
Zeeman-splitted levels as final states are possible. This produces the three-fold splitting of the conductance steps as seen to agree well with experiment in figure~\ref{4}.

\begin{figure*}[htb]
\centering
{\includegraphics[width=5cm,,angle=-90]{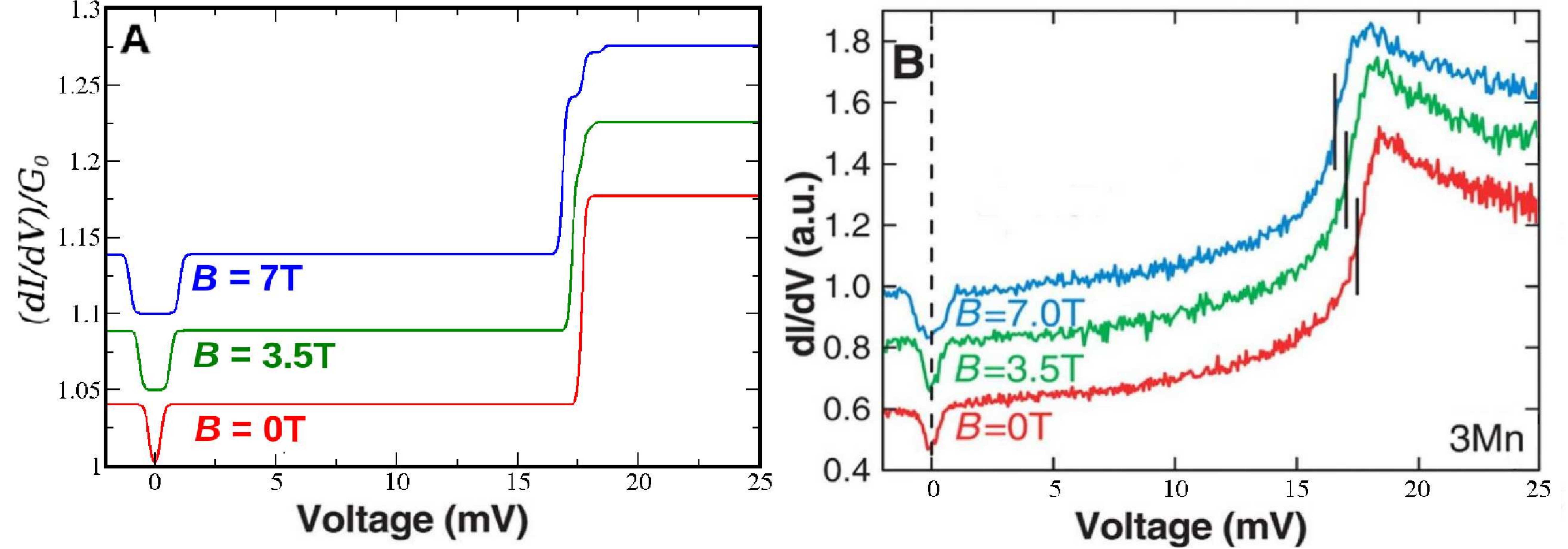}}
\caption{\footnotesize{(Color online) Theoretical (A) and experimental (B) conductance spectra for the $N=3$ trimer at different magnetic field. 
We notice the shift of the first conductance step transition to lower voltages as the magnetic field increases. This is in good agreement with 
STM experiments of reference [\onlinecite{Hir3}], which are reproduced here in panel (B).}}
\label{5}
\end{figure*}

Figure~\ref{5} shows the conductance spectrum for the trimer ($N$=3). This chain exhibits similar trends as those found for the dimer as a 
magnetic field is applied, i.e. there is a Zeeman split of the zero-field conductance steps. From the figure one may notice that for both the 
theoretical (A) and experimental (B) data there is a shift of the principle step to lower voltages as the magnetic field is increased. We remark here that in order to 
recreate the conductance profile of the trimer, additional ferromagnetic second-nearest-neighbour interaction between the local spins at the 
edge of the chain must be included in the model~\cite{Fernandez-Rossier}. The magnitude of this additional exchange parameter is approximately 
half of that of $J_\mathrm{dd}$. The inclusion of such a second-nearest-neighbour coupling constant changes the position of the conductance step from 
a second excited state at 27~mV to a first excited state at 16~mV. This correction is also included in the calculations for $N$=4 (see Fig. \ref{3}), again 
giving good agreement with experiments.

In all of the spectra investigated the most striking agreement with the experiments concerns the correct prediction of the SF-IETS excitation voltages.
In particular not all the possible spin excitations feature in the $\mathrm{d}I/\mathrm{d}V-V$ curve as a result of spin-selection rules. A careful analysis 
of the inelastic process, as outlined in the first section, reveals why some excitations occur and why some other do not. In particular we note that the proper 
selection rules for transitions are governed by the prefactor in the self-energy $|{\langle}m|{S^{i}}|n{\rangle}|^2$ [Eq.~(\ref{eq:15})]. This effectively 
selects which excitations are more probable to occur. For example, the full energy spectrum of the trimer has $6^3=216$ eigenvalues but only a 
small portion of these eigenvalues will contribute significantly to the scattering self-energy. Consequently only a few transitions will have influence on the 
conductance spectrum. This for instance explains why the first conductance step of the dimer is considerably larger than that of other chains of 
different lengths. In fact our calculations show that a transition from the singlet ground state ($S$=0) to any of the triplet excited states 
($S$=1; $m$= -1, 0, +1) has equal probability. This results in a conductance step which is approximately three times larger than any single spin-flip 
event that occurs in an odd chain.

\section{Conclusions}
In conclusion we have formulated a theory of electron transport through magnetic nanostructures based on the NEGF formalism and including inelastic 
spin-scattering. We have first explicitly derived an expression for the interacting self-energy at the level of the first Born approximation in the electron-spin
interaction. This has been used to calculate the current-voltage and the conductance-voltage curves for a 1D system of magnetically coupled 
spins at finite bias and in a magnetic field. Our results reproduce well the features of recent SF-IETS experiments for 1D Mn chains. Most notably 
the severe dependence of the conductance spectra on the chain parity and the selection-rule suppression of certain transitions is a direct outcome
of the theory. Our NEGF approach is therefore a valid alternative to master equation based schemes, with the advantage that it can be scaled up to
larger systems and combined with more sophisticated electronic structure theories.
\section{Acknowledgments}
This work is sponsored by the Irish Research Council for Science, Engineering \& Technology (IRCSET). Computational resources have been provided by the Trinity Centre for High Performance Computing (TCHPC).

\end{document}